\documentclass[review,3p]{elsarticle}

\usepackage[utf8]{inputenc}    
\usepackage[T1]{fontenc} 
\usepackage[english]{babel} 
\usepackage{graphicx}
\usepackage{enumitem} 
\usepackage{hyperref} 
\usepackage{upgreek}
\usepackage{siunitx}
\usepackage{xcolor}
\usepackage{float}
\usepackage{setspace}

\journal{Nuclear Instruments and Methods A}

\begin{document}

\begin{frontmatter}
  \title{XRF element localization with a triple GEM detector using resistive charge division} 
  \author{Geovane G. A. de Souza\corref{corauthor}}
  \cortext[corauthor]{Corresponding author}
  \ead{geovane.souza@usp.br}
  \author{Hugo Natal da Luz}
  \address{Instituto de F\'isica, Universidade de S\~ao Paulo\\Rua do Matão 1371, 05508-090 Cidade Universitária, São Paulo, Brasil}
  
  \begin{abstract}
    In this work we show the operation and results of an X-ray fluorescence imaging system using a cascade of three gas electron multipliers (GEM) and a pinhole assembly. The detector operates in Ar/CO$_2$~(90/10) at atmospheric pressure, with resistive chains applied to the strip readout, which allow to use only five electronic channels: two for each dimension and a fifth for energy and trigger. The corrections applied to the energy spectra to compensate for small changes in the signal amplitude and also differences in gain throughout the sensitive area are described and the clear improvement of the energy resolution is shown.
    To take advantage of the simultaneous sensitivity to the energy and to the position of interaction, a color scale matching the energy spectrum to the RGB range was applied, resulting in images where the color has a direct correspondence to the energy in each pixel and the intensity is reflected by the brightness of the image. The results obtained with four different color pigments are shown. 
  \end{abstract}

\end{frontmatter}

\section{Introduction}

The Gas Electron Multiplier (GEM)~\cite{Sau97} is a Micropattern Gaseous Detector (MPGD) that has undergone a steady development and maturity process over the last two decades. Thanks to its well studied properties such as high counting rate capability, fair energy resolution, good ion backflow suppression and good stability against electrical discharges, allied to the possibility of building detection areas of the order of the square meter, it has been selected to operate in major Particle Physics experiments, such as LHCb~\cite{Car12} and COMPASS~\cite{Alt02} and its use is also foreseen for upgrades in ALICE~\cite{ALICEUP} and CMS~\cite{CMS15}. 

It consists on a kapton foil, typically \SI{50}{\micro\meter} thick, coated on both sides with copper layers. A triangular matrix of biconical holes is etched through the copper and the foil (\SI{70}{\micro\meter} in the copper layer and \SI{60}{\micro\meter} diameter in the kapton substrate). The centers of the holes are at a distance of \SI{140}{\micro\meter} from their nearest neighbors. When the GEM foil is immersed in a gas mixture and suitable voltage differences are applied between two electrodes to define the electric fields above, below and inside the holes of the GEM, the very high electric field inside the holes focuses free electrons generated by the interaction of ionizing radiation with the atoms of the gas. When these electrons penetrate the holes, Townsend avalanches are generated multiplying the primary charge. The possibility of cascading several GEMs, where each GEM multiplies the charge from the preceding one increases significantly the gain in charge of the detector, as well as its stability.

Applications in low energy Physics, namely in X-ray imaging have been developed towards very high resolutions, often at a cost of the detection areas. In fact, the newest room temperature solid state detectors achieve remarkable energy and position resolutions in the \si{\micro\meter} range. This high resolution spans small areas that range a few \si{\square\cm} at most. In the study of historical artifacts or art pieces by X-ray Fluorescence, usually areas in the order of hundreds of \si{\square\cm} must be studied, with a resolution around 1\,mm. This is usually done with small solid state detectors without position sensibility, that are scanned through the whole area that must be studied. GEM-based detectors can present an advantage, reconstructing elemental distributions over large areas in a much shorter time, sparing the need of long and tedious scans. This approach has been gaining some space, with different groups using different MPGD and presenting promising results with GEM~\cite{Zie13}, Micro-Hole \& Strip Plate (MHSP)~\cite{Sil11} and Thick-Cobra\cite{Sil13}. Reference~\cite{Vel18} makes a review on some of the work done by different research groups using MPGD with this technique.

The energy resolution limitations of this type of detector with respect to the solid state ones is obvious. Nevertheless, by applying corrections related to variations of the signal amplitude during acquisitions and also to local gain variations throughout the sensitive areas improve very much its performance, promising a very valuable tool in X-ray Fluorescence Imaging.

\section{Experimental setup}

\begin{figure}[h]
  \centering
  \includegraphics[width=7.5cm]{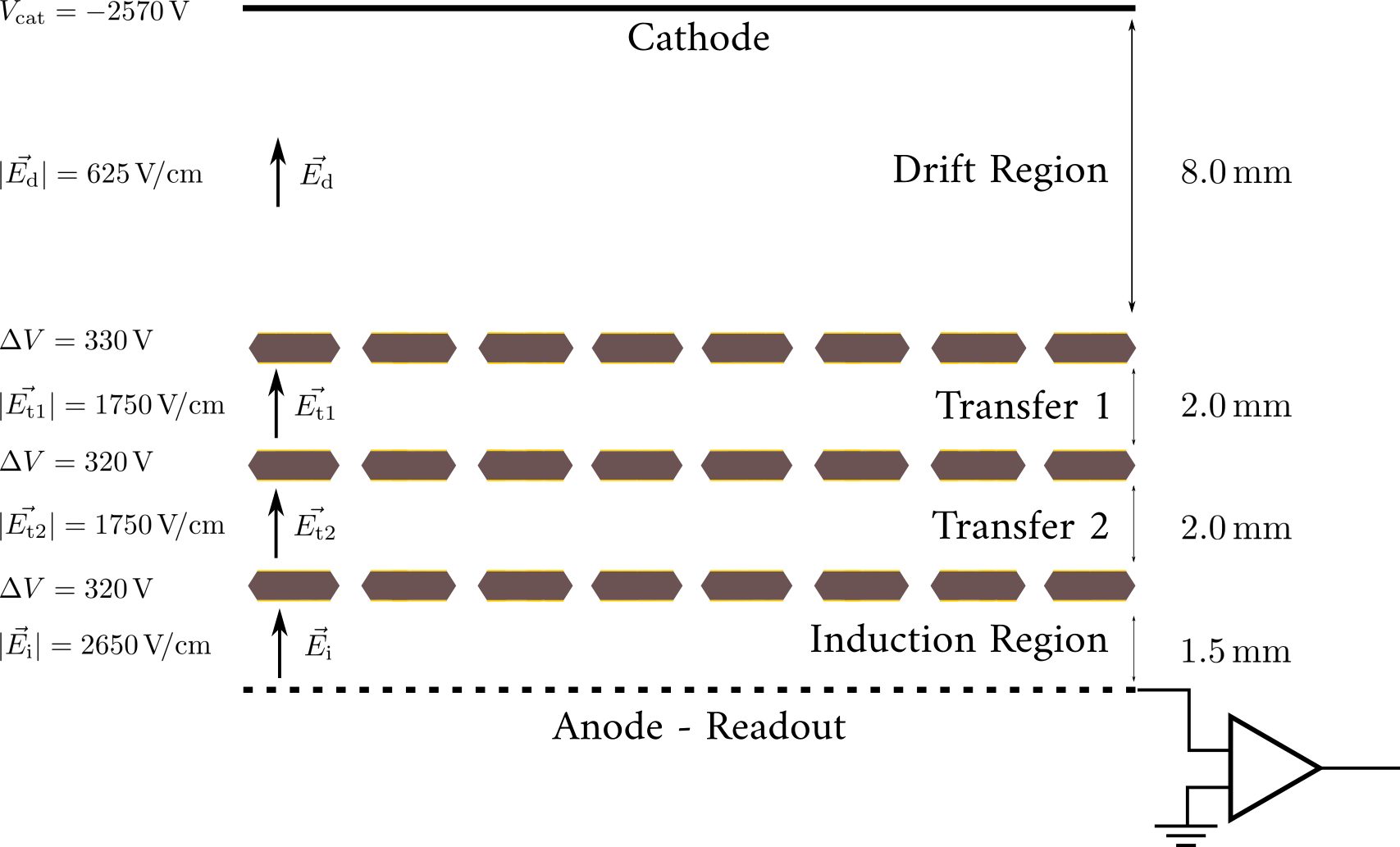}
  \caption{The triple-GEM setup.}
  \label{ImaGEM2}
\end{figure}

The detector consists on a cascade of three GEMs immersed in a mixture of Ar/CO$_2$\,(90/10) at atmospheric pressure. The detector window consists of a kapton foil \SI{50}{\micro\meter} thick. The gas passes through \SI{4}{\milli\meter} tubes and the flow is set to 6 l/h. The triple-GEM geometry can be seen in figure~\ref{ImaGEM2}, where the dimensions and typical electric fields and voltages are also depicted. 

\begin{figure}[h]
  \centering
  \includegraphics[width=7cm]{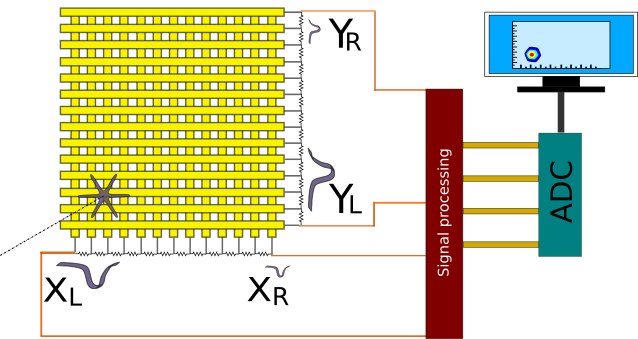}
  \caption{Scheme of the strip readout system using resistive chains.}
  \label{readout}
\end{figure}

The hole pitch of the GEMs is \SI{140}{\micro\meter}. The readout system is segmented in 256 strips in each dimension~(fig.\ref{readout}), which are interconnected through resistive chains. By collecting the charge at both ends of each resistive chain, it is possible to calculate the projection of the primary X-ray ionization on the X--Y plane for each coordinate through eq.~\ref{interaction}. 

\begin{equation}
  x= l \frac{X_L-X_R}{A},\qquad
  y = l \frac{Y_L-Y_R}{A}
  \label{interaction}
\end{equation}

\noindent where $X_L$, $X_R$, $Y_L$ and $Y_R$ are the signal amplitudes for the left and right ends of the $X$ and $Y$ resistive chains according to figure~\ref{readout}, $l$ is the length/width of the detector and $A$ is given either by the sum of the amplitudes of all four channels or by the amplitude of the signal collected from the bottom electrode of the last GEM. This signal also served as the global trigger of the electronic system. 

The charge collected by each channel is integrated by a standard charge sensitive pre-amplifier with a charge sensitivity around 1\,V/pC and a rise time around 50\,ns, and shaped by differentiating the signals, resulting in a gaussian peak with a width of around $\SI{3}{\micro\s}$, which is suitable for the counting rates used throughout this work. After application of simple logic, it is sampled by a standard 12 bit peaking ADC.

As X-ray source we used the Amptek Mini-X~\cite{minix} with a silver target, operating at a high voltage of typically 15\,kV and filament current around 15\,$\upmu$A. We also used a $^{55}$Fe radioactive source, which decays into manganese by electron capture emitting 5.9\,keV (K$_{\alpha}$) and 6.4\,keV (K$_{\beta}$) characteristic X-rays, for energy calibration and to determine the energy resolution. A framework for data processing, image reconstruction and analysis was developed using the ROOT framework developed by CERN~\cite{root} and other C++ libraries.

To characterize the detector in terms of intrinsic position resolution, different masks were imaged in transmission mode, with the object placed between the detector and the X-ray source, directly on the detector window, ensuring that there was no magnification in the transmission images obtained.

\begin{figure}[h]
  \centering
  \includegraphics[width=0.4\textwidth]{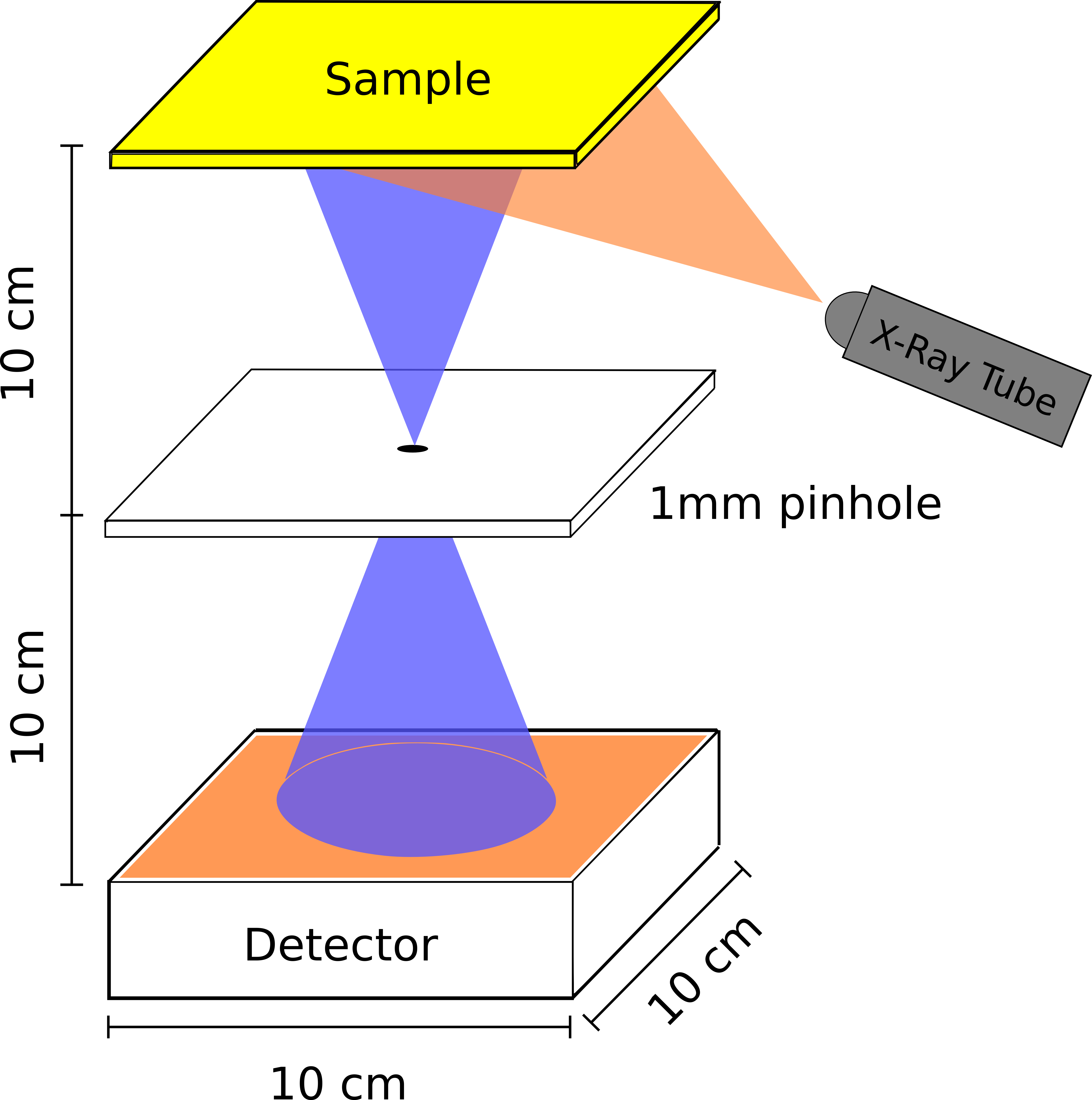}
  \caption{The pinhole setup used in this work for X-ray fluorescence imaging.}
  \label{fig:pinhole}
\end{figure}

For the X-ray fluorescence imaging, a stainless steel pinhole with diameter and thickness of 1mm was placed between the sample and the detector window, at 10\,cm from both, leading to a magnification of 1. The sample object was irradiated with the high intensity X-ray source. The fluorescence X-rays crossed the pinhole before entering the detector, as shown in figure~\ref{fig:pinhole}. The pinhole assures that the photons arriving at each point in the sensitive area of the detector correspond univocally to one position in the sample, thus allowing to accurately reconstruct the elemental distribution in the sample. To test this capability, a set of four different pigments was irradiated and imaged. 

\section{Results and discussion}
\label{sec:res}

\subsection{Energy resolution --- signal amplitude corrections}
\label{sec:corr}
The energy resolution was measured using a $^{55}$Fe radioactive source. The full window area was irradiated with the source, which was placed about 10\,cm above the detector window. During the irradiation time, which in the final application is expected to be 3 to 4 hours long, small drifts in the detector's gain and consequently in the amplitude of the signals are expected reflecting the changes in environmental conditions of the detector surroundings, such as the temperature or atmospheric pressure. The study of the mechanisms of this influence is beyond the scope of this work, however, gain changes were forced by changing the temperature of the room and test the algorithms for corrections of temporal variations of the signal amplitude.

\begin{figure}[h]
  \centering
  \includegraphics[width=0.5\textwidth]{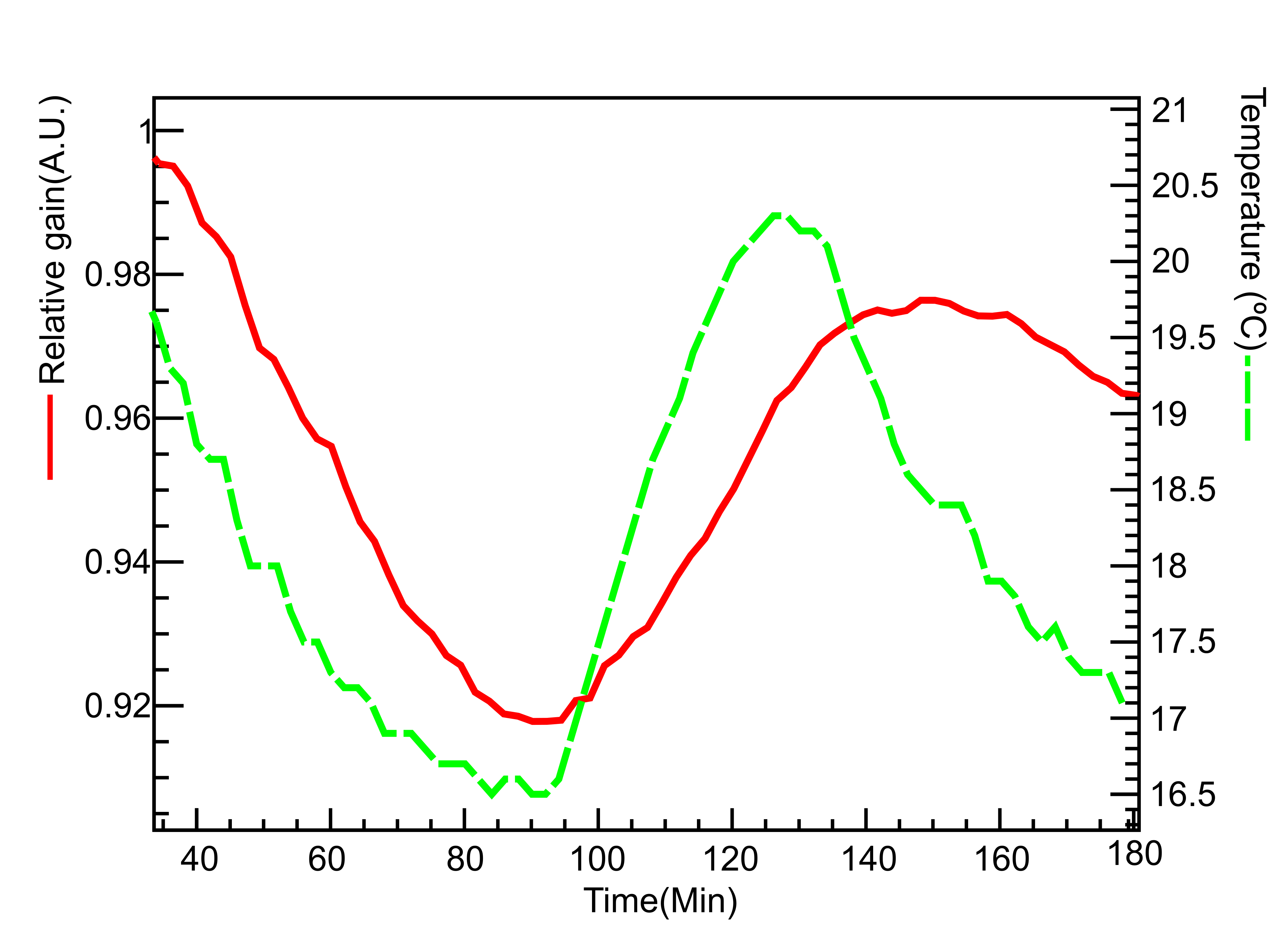}
  \caption{Detector's gain and room temperature measured over three hours, while varying the temperature of the lab.}
  \label{gaintemp}
\end{figure}

Figure~\ref{gaintemp} shows how a change in the gain of the detector can be induced by changing the environment temperature.

\begin{figure}[h]
  \centering
  \includegraphics[width=0.48\textwidth]{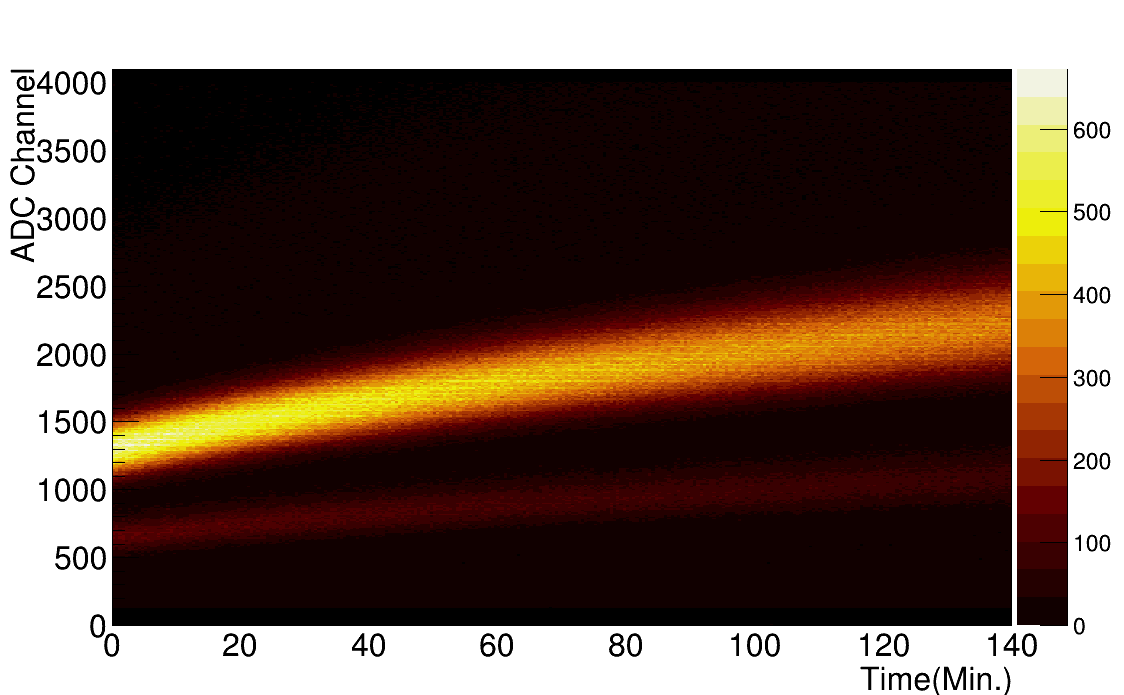}    \includegraphics[width=0.48\textwidth]{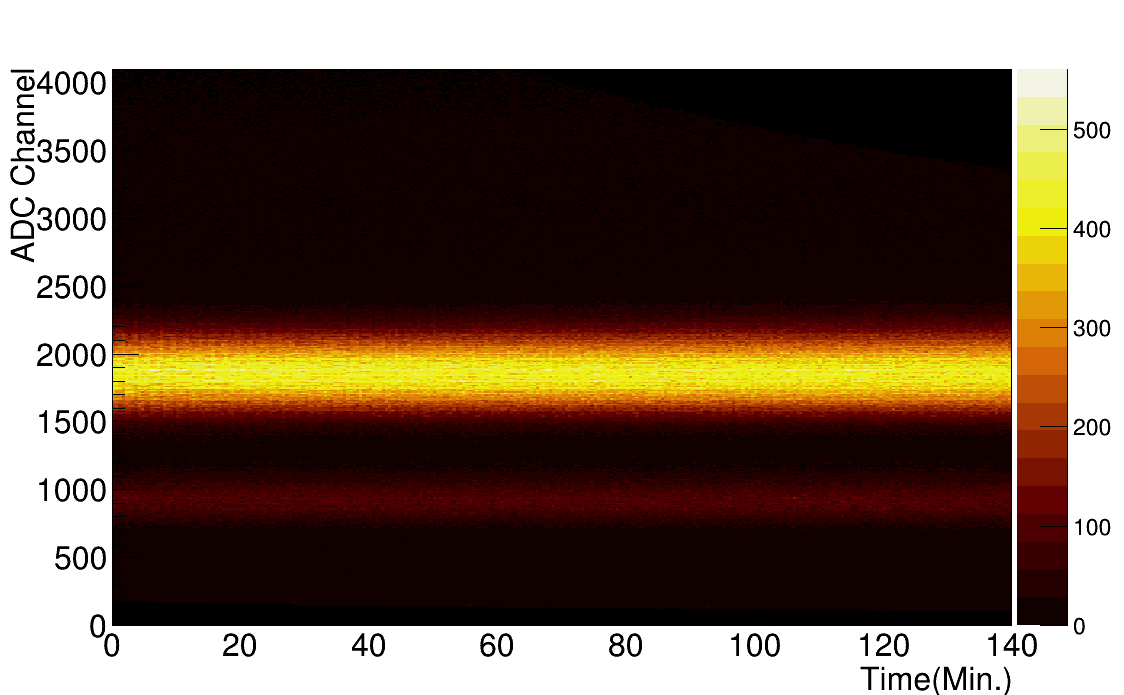}
  \caption{Left: Distribution of the signal amplitude obtained with the $^{55}$Fe radioactive source over time. The color scale indicates the number of counts. The bright part of the spectrum is the main peak of the energy spectrum. In this example, the amplitude of the signals increased slowly during 140 minutes. Right: The distribution on the top, with the signal amplitude corrected. The variations in the amplitude over time were corrected for the whole energy spectrum, including adjacent peaks, such as the scape peak.}
  \label{1stcorrection}
\end{figure}

The correction applied to the signal amplitude was done offline and consisted in dividing the acquired data in time slices, each one containing enough events to allow a correct determination of the center of the main peak. A number of $10^5$ events in each time slice was considered more than sufficient for this correction and can be reduced in case of large and fast changes. After this, the position of the main peak is normalized between all the time slices, eliminating the temporal differences of the signal amplitudes, as shown in figure~\ref{1stcorrection}. This correction can be done to every set of data, regardless of the reason that made the signal amplitudes drift.

\begin{figure}[h]
  \centering
  \includegraphics[width=0.48\textwidth]{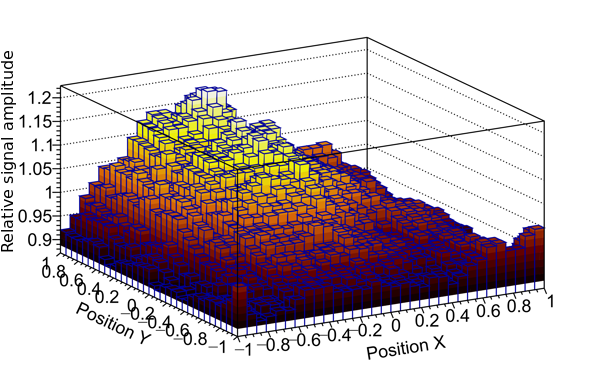}    \includegraphics[width=0.48\textwidth]{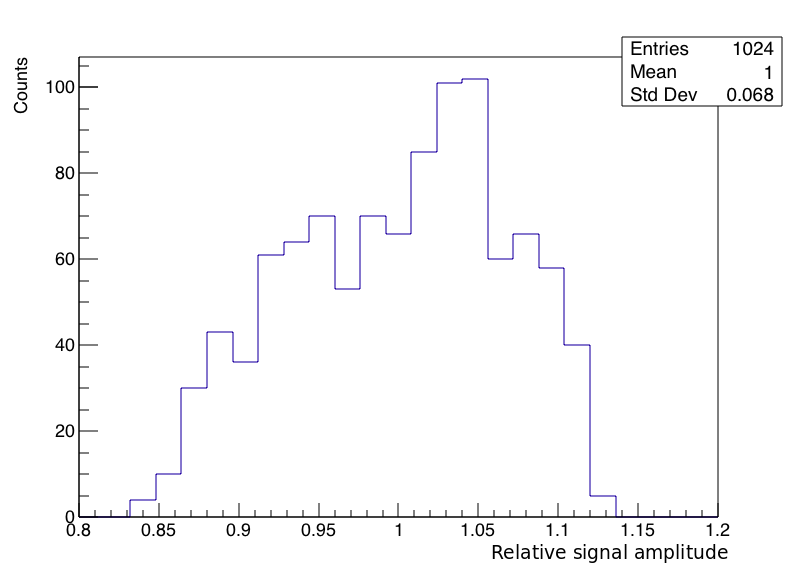}
  \caption{Left: the amplitude of the signals across the detector normalized to the average. Right: the amplitudes show a standard deviation of 7\,\% throughout the sensitive area of the detector.}
  \label{mapa}
\end{figure}

Besides the corrections due to drifts of the amplitude in time, a second offline correction was done to compensate for the gain non-uniformity across the detector area. It is know that many factors such differences of the dielectric thickness, different hole diameters or even slight staggering of the GEM foils may locally affect the detector gain and, consequently, the energy resolution. To overcome this issue the effective area of detection was divided into 1024 different sectors and, again, the center of the main peak was determined. The correction factor is calculated by dividing the correspondent position of the peak in ADC channels by the average of all 1024 sectors. Figure~\ref{mapa} shows the main peak position relative to the average of the 1024 sectors. The gain variations have a standard deviation of 7\,\% and could be as high as 20\%.

\begin{figure}[h]
  \centering
  \includegraphics[width=0.48\textwidth]{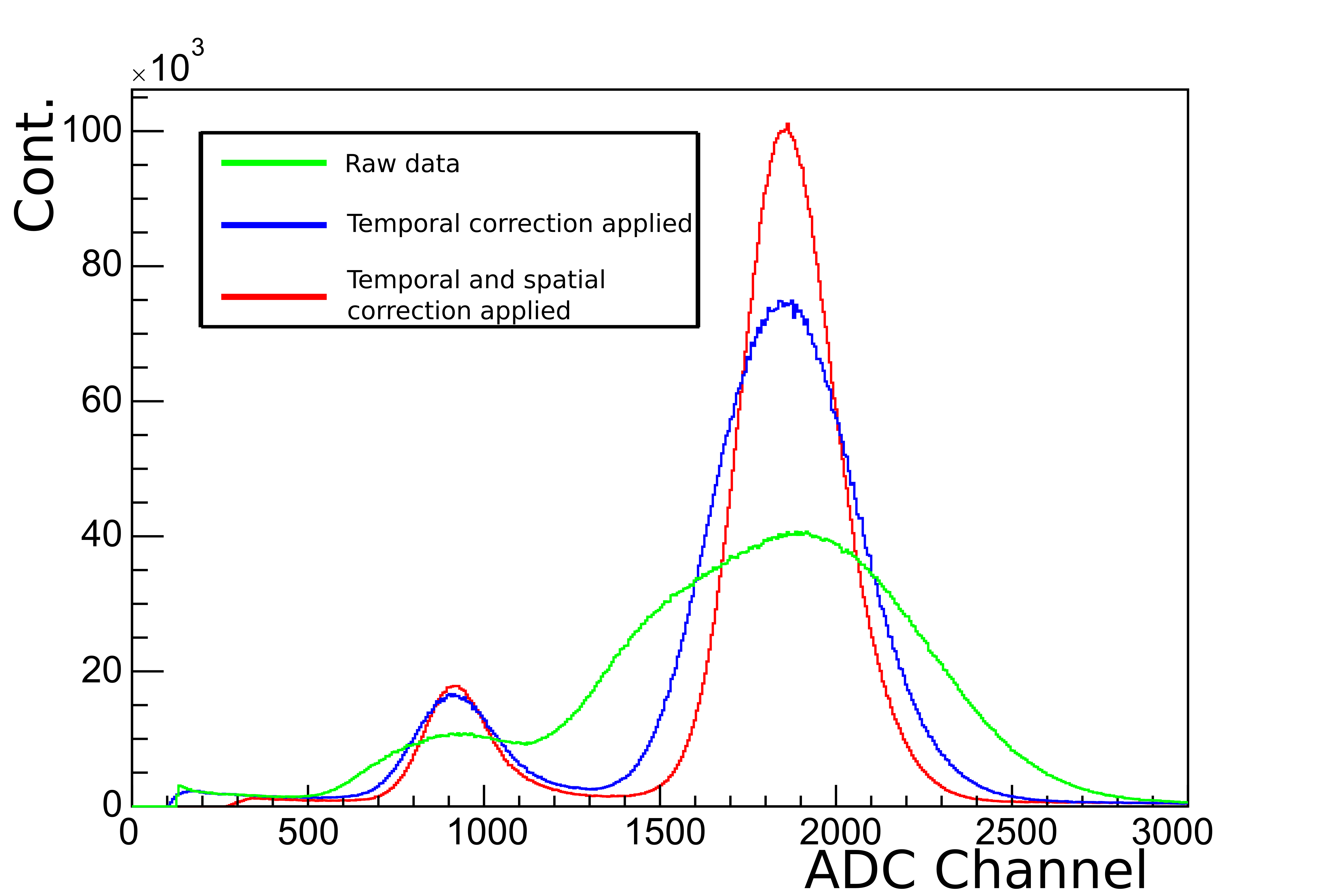}
  \includegraphics[width=0.48\textwidth]{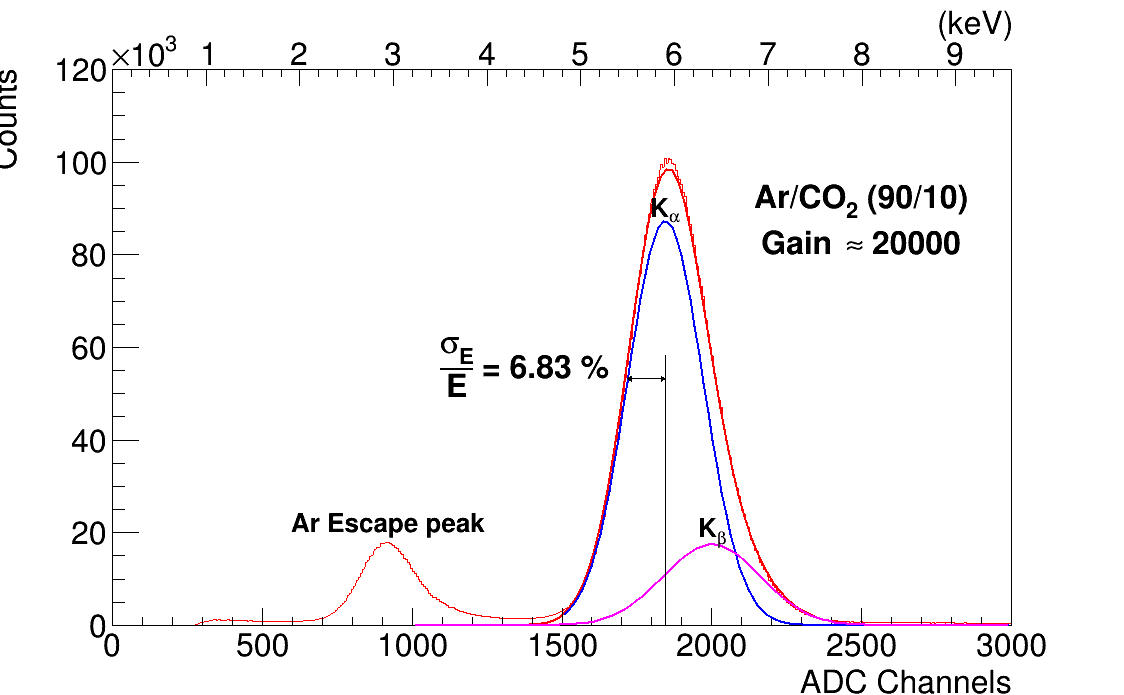}
  \caption{Left: Green --- The raw energy spectrum; Blue --- after temporal correction; Red --- both temporal and spatial corrections applied. Right: The energy resolution obtained after the corrections is 6.8\%~($\sigma$).}
  \label{X}
\end{figure}

The spectra before each step of correction and the final one can be seen in figure~\ref{X}. As we can see, the energy resolution improves significantly with each correction, achieving in the end 6.8\% energy resolution~($\sigma$), after fitting the two gaussian curves corresponding to the K$_\alpha$ and K$_\beta$ lines of manganese. The argon escape peak can also be seen at around 2.9\,keV in the energy distribution. This peak is related to argon fluorescence X-rays from the K shell that escape the detector, a well-known effect (see~\cite{escape} for example). The presence of this escape peak is inevitable in detectors using argon mixtures and will be discussed further ahead.

\subsection{Position resolution}

\begin{figure}
  \centering
  \includegraphics[width=0.32\textwidth]{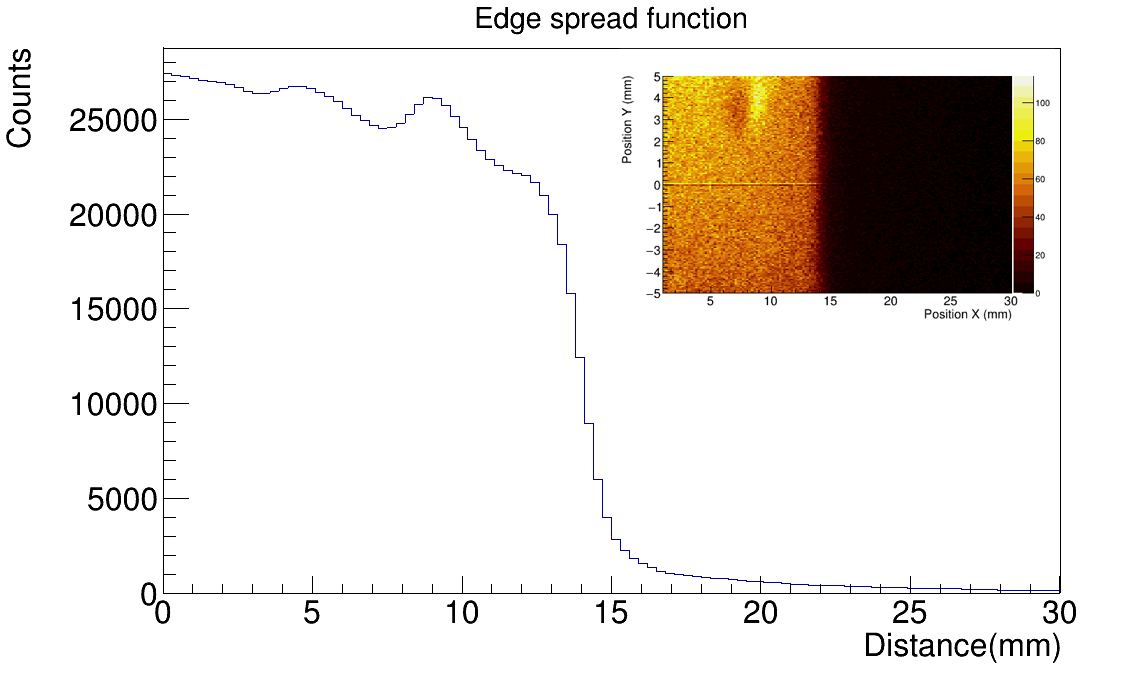}  \includegraphics[width=0.32\textwidth]{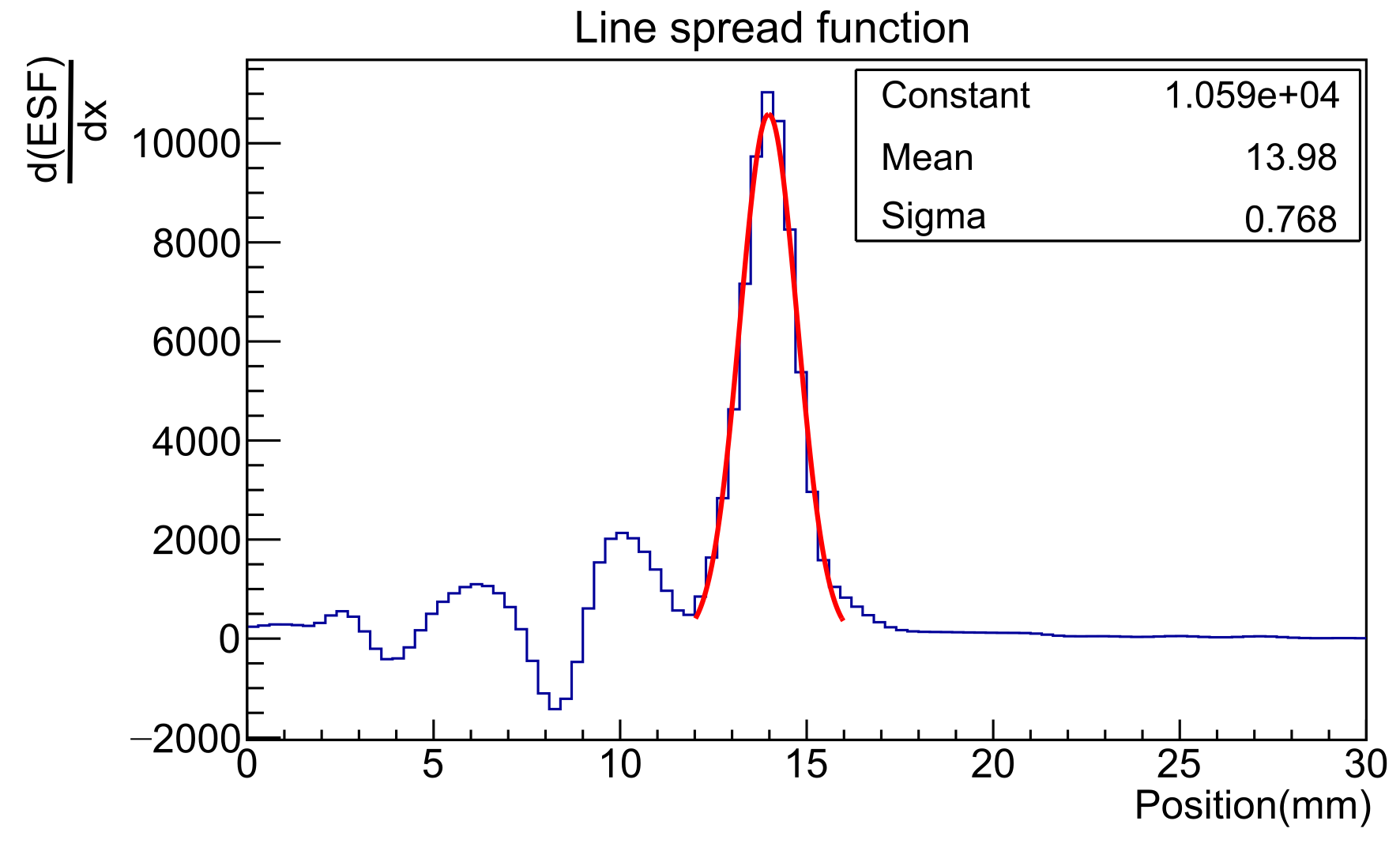}
  \includegraphics[width=0.32\textwidth]{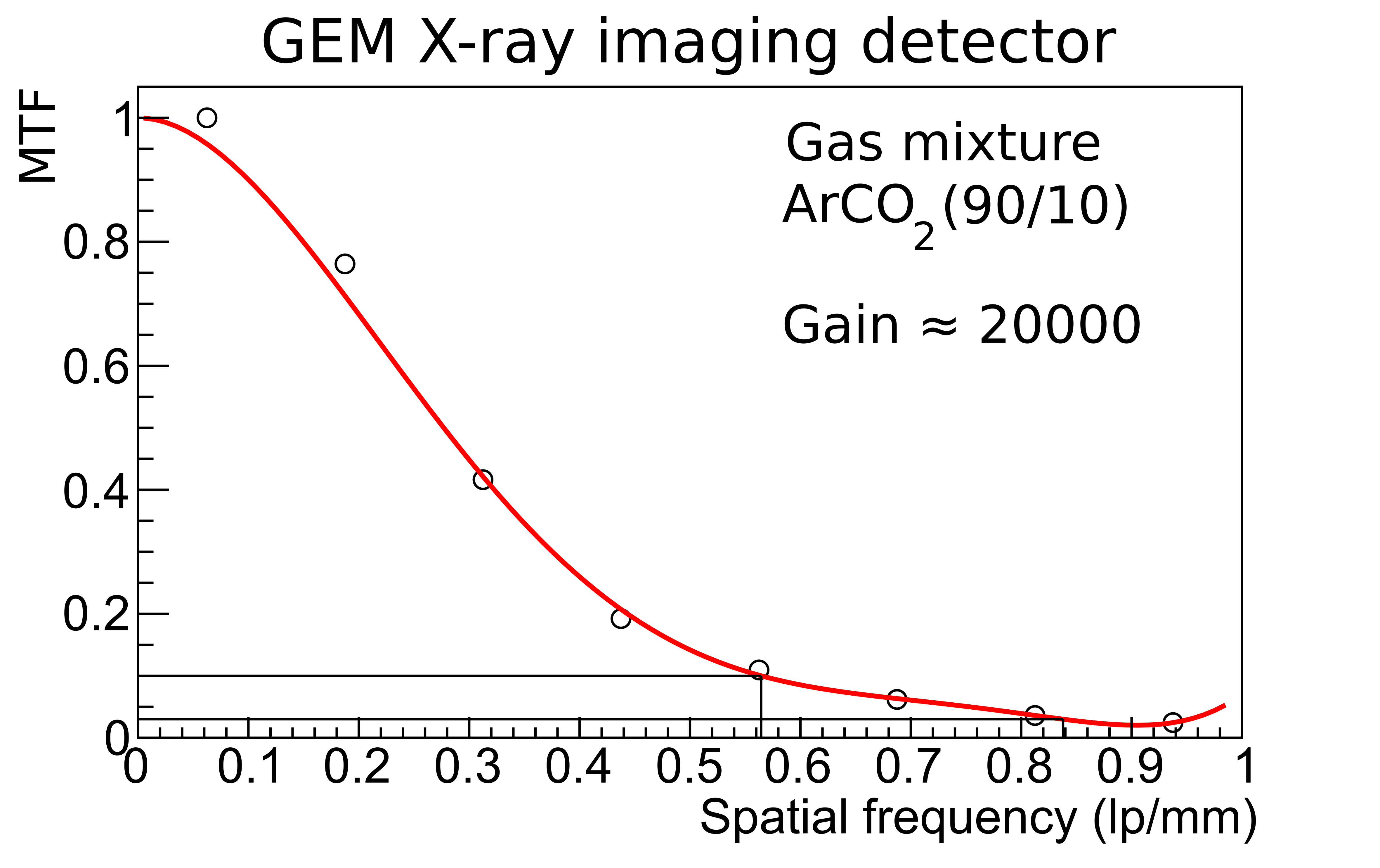}
  \caption{The Modulation Transfer Function obtained by imaging a sharp edge. Left: a sharp edge profile in the image is the Edge Spread Function (ESF). The ESF is derivated resulting in the Line Spread Function (LSF). Right: The Fast Fourier Transform of the LSF results in the Modulation Transfer Function (MTF). The red curve is not a fit, it is placed to guide the eye. The position resolution is usually taken from the MTF at 10\,\% (marked in the plot), which is around 1.8\,mm($\frac{1}{0.56}$\,lp/mm), consistent with the width of the LSF.}
  \label{fig:MTF}
\end{figure}

The imaging capability of the detector was characterized using X-ray transmission through different masks that allowed to calculate its performance in terms of resolution and contrast as a function of the spatial resolution. One of the most reliable methods for estimating the performance of an imaging system is calculated from the image of a sharp edge. The edge intensity profile is a step function, where its slope is a measure of the position resolution. The width of its derivative is used to quantify this slope and by application of a Fourier Transform, the contrast of the imaging system as a function of the spacial frequencies, i.e., the Modulation Transfer Function (MTF) can be estimated. Figure~\ref{fig:MTF} shows process of obtaining the MTF obtained for this detector, when the full energy spectrum from a silver target X-ray source was used. The distances in the image were calibrated by using the known distance between the two edges. 

\begin{figure}
  \centering
  \includegraphics[width=0.48\textwidth]{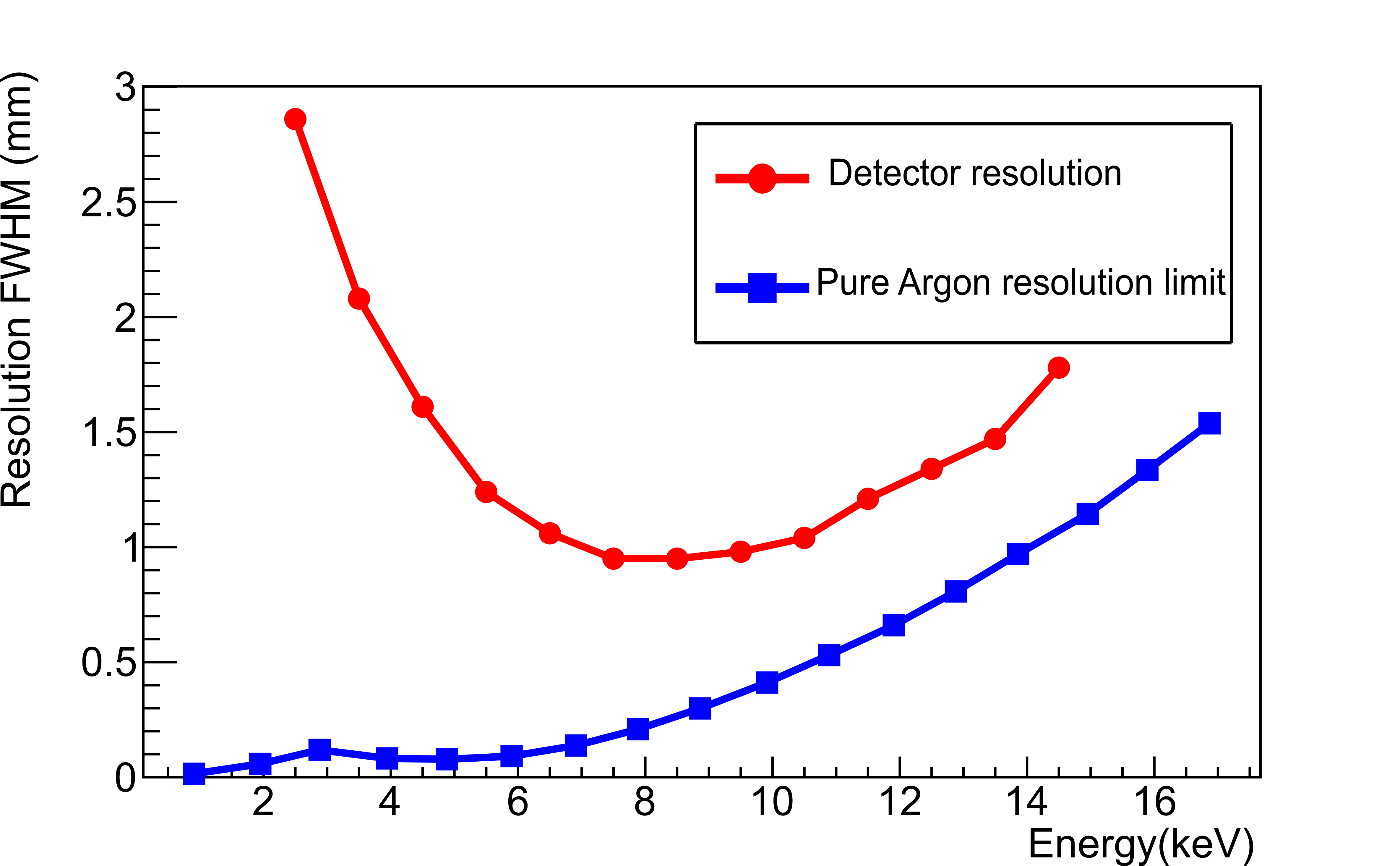}
  \caption{The position resolution of the detector as a function of the X-ray energy (red line). The position resolution worsens for higher energies due to the higher range of the photo-electrons. For the lower energy, its deterioration is related with the poorer signal-to-noise ratio. The blue line indicates the resolution limit for pure Argon as simulated in~\cite{Aze15}.}
  \label{fig:range}
\end{figure}

One of the features of this type of detector is the dependence of the spatial resolution on the X-ray energy. Figure~\ref{fig:range} shows that dependency in this specific detector. It is directly related to the range of the photo-electrons, that in Argon, above the K-absorption edge at 3\,keV, increases monotonically. Since the data was collected event-by-event, it was possible to measure resolutions of images only for certain energy range. With this, it was possible to study the dependence of the position resolution with the energy. For lower energies the position resolution also worsens due to the smaller signal-to-noise ratio. This has been studied for this detector and reported in~\cite{EXRS}. The best result in terms of position resolution is 1.2\,mm, achieved for the 8--9\,keV range.

\begin{figure}
  \centering
  \includegraphics[height=5cm]{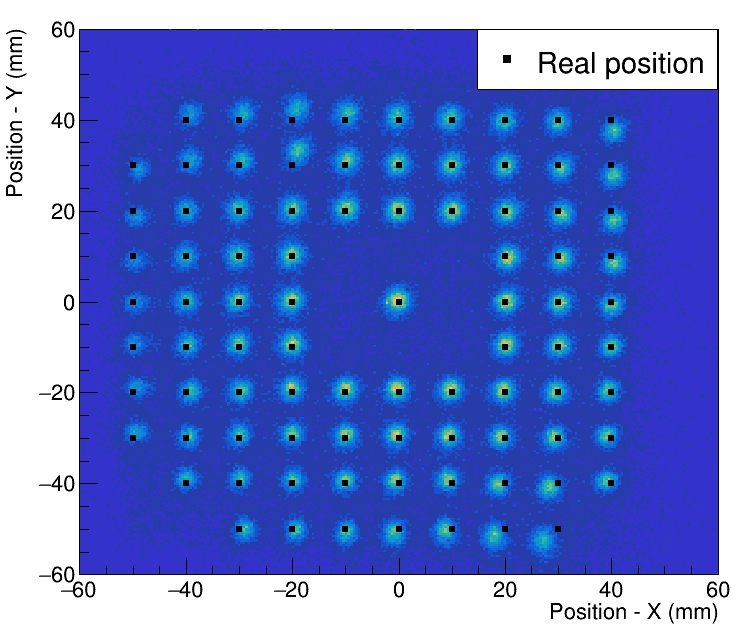}  \includegraphics[height=5cm]{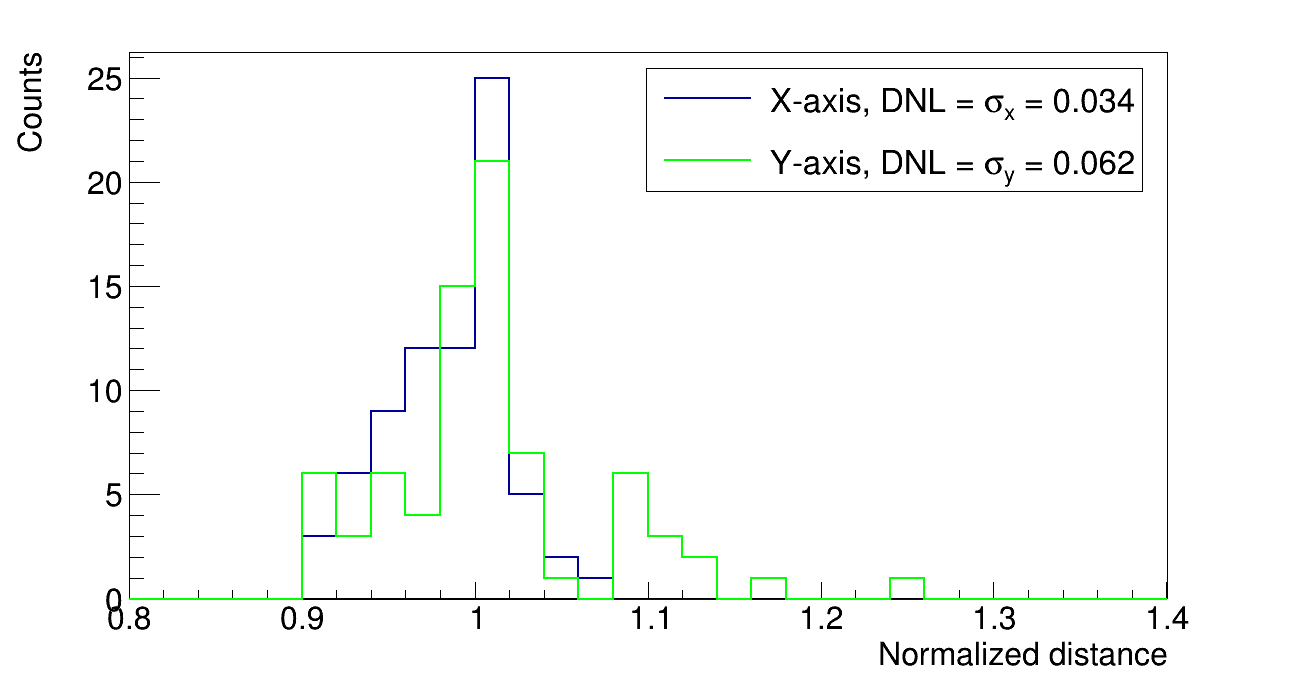}
  \caption{Left: the image of a square matrix of 1\,mm holes with a pitch of 10\,mm. Right: Histogram of the distance between each pair of holes as measured by the detector, normalized by the real distance of 10\,mm. The standard deviation of these distributions is the differential non-linearity which is 3.4\,\% and 6.2\,\% for $x$ and $y$, respectively.}
  \label{fig:DNL}
\end{figure}

To address the image distortions caused by the resistive chain either due to resistor inaccuracies or to imperfections in the mounting of the components, not only the resistors, but also the 128 pin connector, a study of the differential non-linearity was carried out. The differential non-linearity is defined here in analogy to the analog to digital converters, and is the deviation between the distances measured by the detector, relatively to their correct value. For this, a 1\,mm thick steel plate perforated with a square matrix of 1\,mm holes at a pitch of 10\,mm was placed on the detector window and imaged with the X-ray source at a reasonable distance to keep the magnification negligible. Figure~\ref{fig:DNL} shows the imaged plate on the left, with the real position of the holes marked with black dots. A histogram of the distance between the pairs of holes in the image, normalized by the real one (10\,mm) is shown in the left side of fig.~\ref{fig:DNL} for the $x$ and $y$ directions. The DNL is given by the standard deviations of the distributions and is 3.4\,\% and 6.2\,\% for $x$ and $y$, respectively.

\subsection{X-Ray fluorescence imaging}

\begin{figure}[h]
  \centering
  \includegraphics[width=0.45\textwidth]{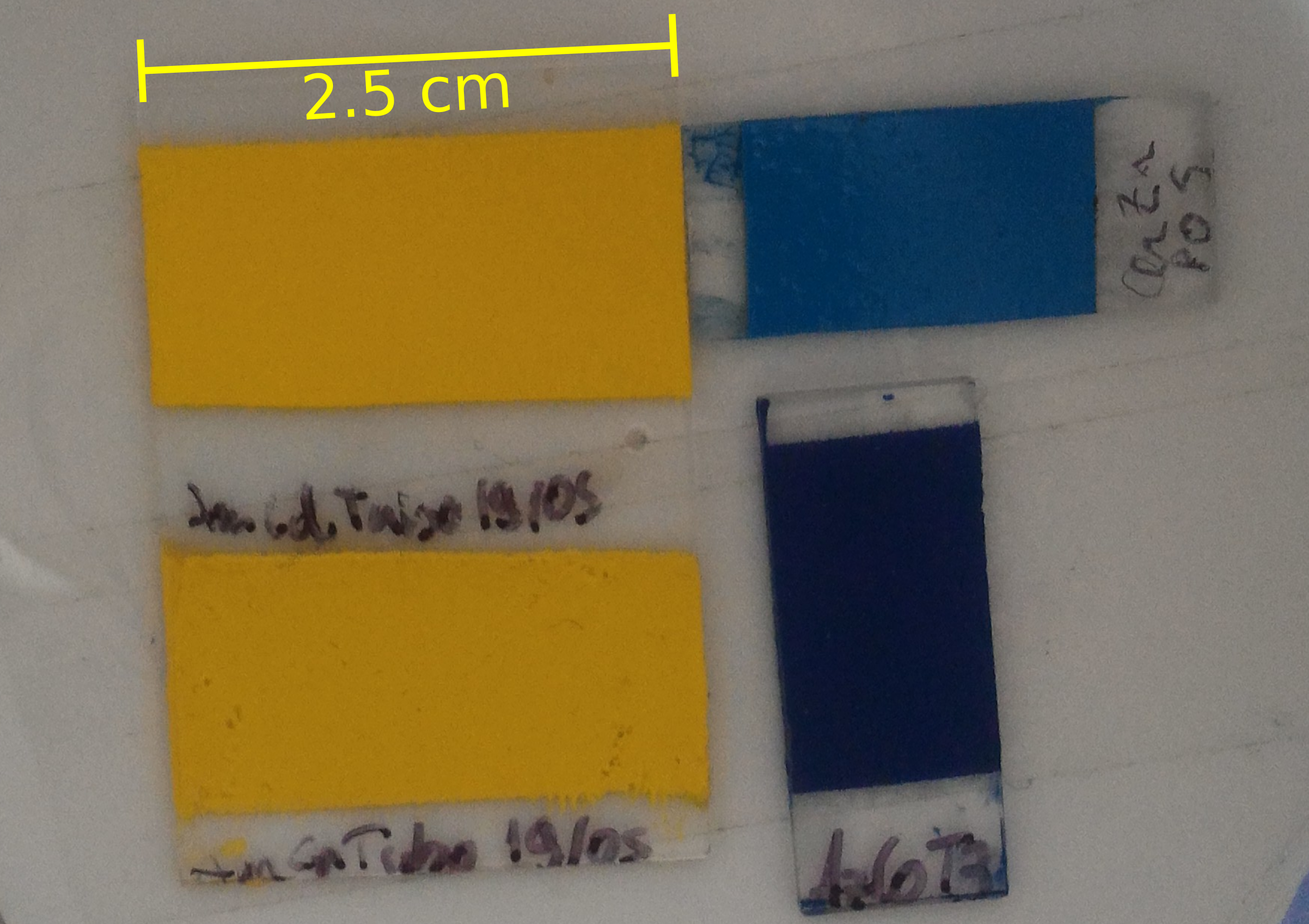}\\  
  \caption{The pigments used to test the X-ray Fluorescence Imaging system. Top-left: Cadmium yellow. Bottom-left: Chromium yellow. Top-right: Cerulean blue. Bottom-right: Cobalt blue.}
  \label{fig:pigs}
  \includegraphics[width=0.5\textwidth]{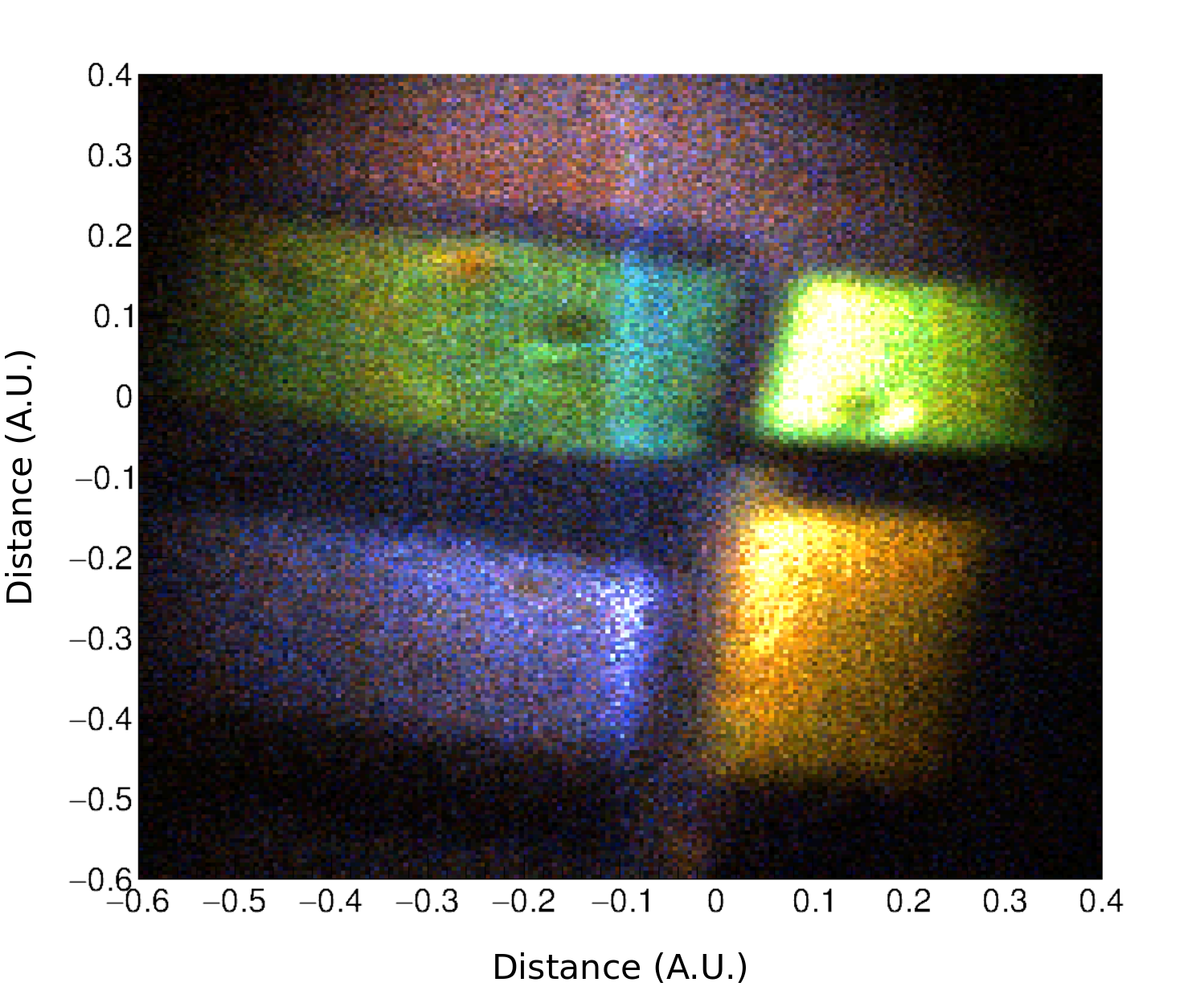}\includegraphics[width=0.45\textwidth]{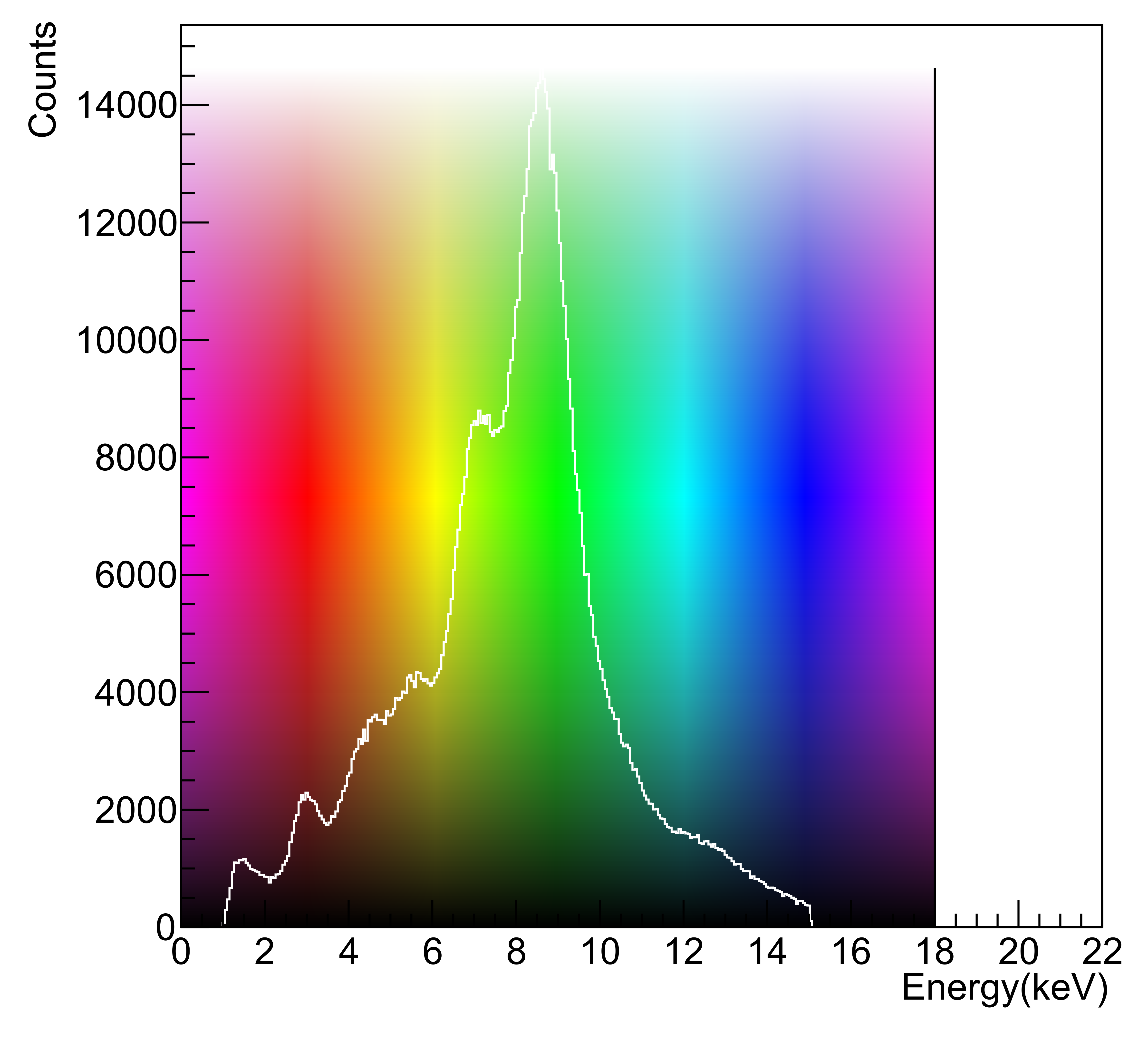}
  \caption{Top: The energy spectrum for the total area. Bottom: XRF image generated for the four pigments. The colors indicate the mean energy deposited in each pixel.}
  \label{fig:XRF}
\end{figure}

The X-ray Fluorescence Imaging capability of the system was tested by irradiating a set of four different ink pigments. The pigments were chrome yellow, cadmium yellow, cerulean blue and cobalt blue. The pigments may also contain zinc white ink in their composition~\cite{pigments}. Figure~\ref{fig:pigs} shows the pigments used. They spanned an area around \SI{50}{\square\cm}. The sample was irradiated during 4 hours with a filament current of \SI{200}{\micro\ampere}.

\begin{figure}[h]
  \centering

\end{figure}

The reconstructed total energy distribution after the corrections described in subsection~\ref{sec:res}.\ref{sec:corr}, and elemental map of the sample is shown in figure~\ref{fig:XRF}. The color scale in the image does not reflect the X-ray intensity, but the average energy of the spectrum in each pixel, as shown by the RGB rainbow (built from the red, green and blue palletes) in the energy spectrum. The X-ray intensity is given by the brilliance. 

\begin{figure}[h]
	\centering
	\includegraphics[width=0.5\textwidth]{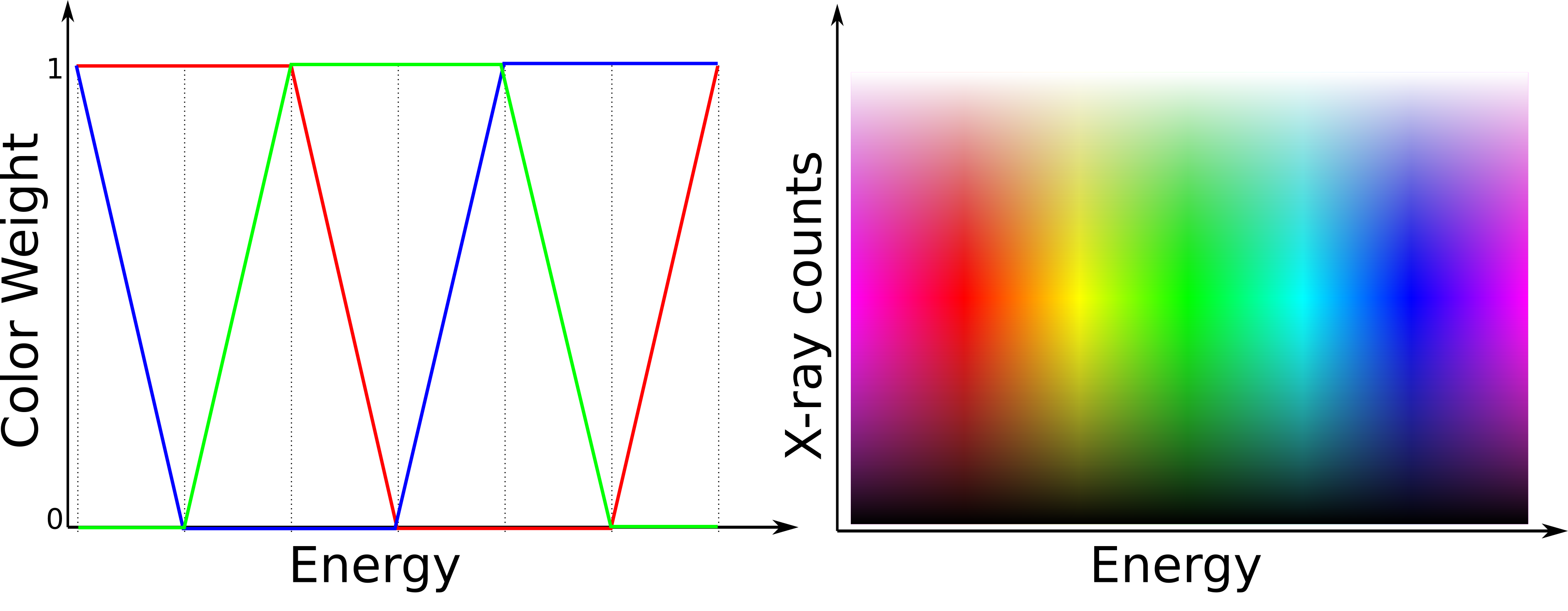}
	\caption{Left: the colors at the limit of the RGB spectrum are adjusted to the maximum and minimum energies of the X-ray energy spectrum. Right: the intensity of each X-ray energy is given by the brightness: more dark for lower intensities and more bright for higher intensities.}
	\label{fig:EscalaCores}
\end{figure}

To match the energies with the color scale, each ADC value is weighted by the correspondent amount of each of the three primary colors as figure~\ref{fig:EscalaCores} shows. Based on this weight, tree different images --- one for each primary color --- are reconstructed. After a trivial leveling of the colors for white balancing, the images are merged, resulting in an image that has information of both the intensity and the energy of the X-rays. This makes an automatic color separation of the different elements in the sample.

\begin{figure}
  \centering
  \includegraphics[width=0.45\textwidth]{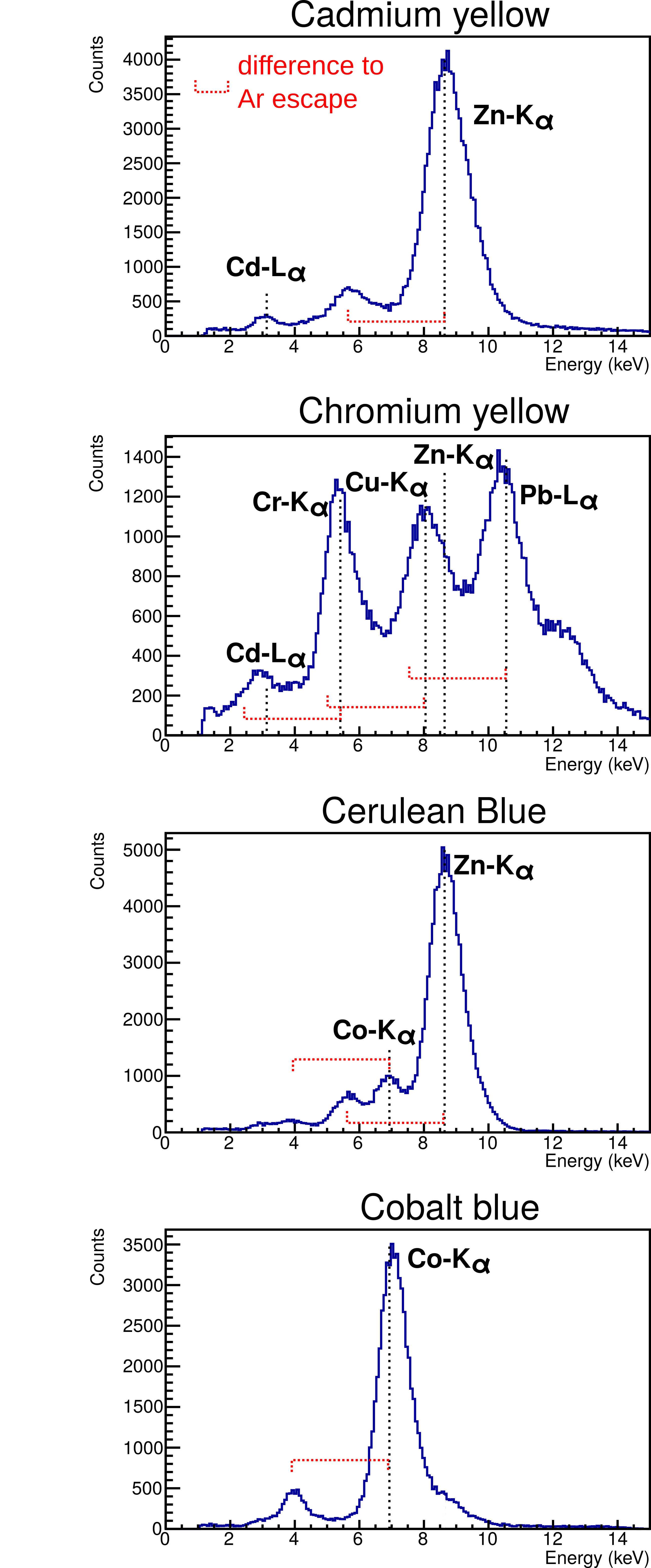}
  \caption{Energy spectrum for each pigment. The dotted black lines show the expected elements for each sample. The dotted red rules mark an energy difference of 3\,keV to each peak, showing the expected position of the argon escape peaks. }
  \label{fig:spectra}
\end{figure}

From the energy spectra of the four regions corresponding to the four pigments, it is possible to extract information of the elements present in each one. The energy distributions are shown in figure~\ref{fig:spectra}, where the main peaks are identified. 

As mentioned before, one feature of this type of detector, using argon-based mixtures is the presence of an argon escape peak associated to every major peak, appearing at an energy 3\,keV smaller. The interpretation of the energy spectra must take this into account. Some of the argon escape peaks are identified in the spectra of figure~\ref{fig:spectra} with a red scale, showing how they can overlap other peaks or the background, making the identification of elements somehow more complex.

The differences between the two spectra of the yellow pigments are very clear. Where the cadmium yellow shows zinc and cadmium in its composition, the chromium yellow pigment contains copper and lead, besides chromium. The leftmost peak at around 3\,keV and the traces of zinc evident by the small shoulder in the copper peak, suggest a small contamination of this pigment with cadmium yellow.
In the blue pigments, it is interesting to note that both were composed of cobalt and zinc, but in clearly different concentrations.

It is important to notice that this analysis is qualitative. In the case of the blue pigments, it is possible to conclude that the cerulean blue has a higher concentration of zinc, when compared with the cobalt blue. However, the determination of the absolute concentration of each one of the elements must take into account the X-ray emission yields and the efficiency of the detector for each X-ray energy range. This efficiency drops for higher energies due to the small thickness of the absorption region, which is 8\,mm. For the lower energies, the kapton window and cathode, with a total thickness of \SI{100}{\micro\m}, and the distance the X-rays must travel in air before entering the detector also limit the efficiency. 

\begin{figure}
  \centering
  \includegraphics[width=0.45\textwidth]{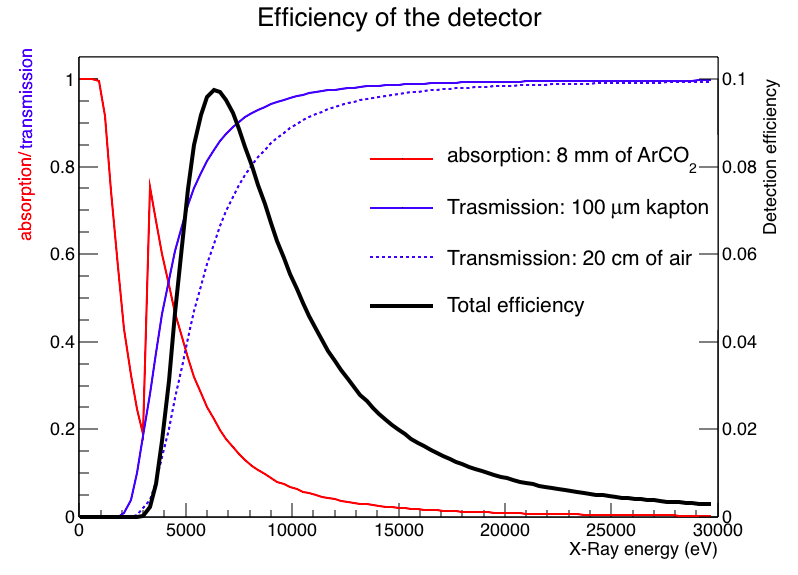}
  \caption{The detection efficiency taking into account the transmission of X-rays through the two layers of kapton in the detector and 20\,cm of air and absorption in the 8\,mm of Ar/CO$_2$ in the drift region. The axis in the left refers to the transmission and absorption of the X-rays (red and blue curves) and the axis on the right refers to the detection efficiency (black curve).}
  \label{fig:efficiency}
\end{figure}

The efficiency calculated by taking into account the fraction of X-ray photons hitting the detector that are actually detected is shown in figure ~\ref{fig:efficiency}. This calculation is done from the absorption coefficients from the argon-based mixture and the transmission of the kapton and the layer of air, according to the geometry of the system, and using the X-ray interaction tables from~\cite{Hen93}. The detection efficiency peaks around 7\,keV and drops very sharply for lower energies. The efficiency for lower energies can be increased by changing the materials of the window and the cathode. For higher energies, the drop in efficiency is less steep. On this side, the efficiency can be increased by increasing the depth of the drift region.

\section{Conclusion and outlook}

The position sensitive gaseous detector prototype for X-ray fluorescence imaging using a pinhole described in this work is capable of mapping elemental distributions in space with a position resolution close to 1\,mm. Its large sensitive area (\SI{100}{\square\cm}) and simultaneous sensitivity to both energy and position make it suitable to study large distributions, such as those seen in paintings or other cultural and archaeological artifacts. 

The elemental distribution in space is reconstructed with a single data acquisition, without the need of scanning the detector or the X-ray source through the object. The gain in time and system simplicity is an advantage that can make this concept competitive, even with the disadvantages of limited energy and position resolutions. The corrections done to the energy distributions to compensate for small drifts in the gain during the acquisition and for the spacial non-uniformities of the detector kept the energy resolution at the same value as when only a small area of the detector is used. This gives very good perspectives for scaling the sensitive area, to obtain even larger images, eventually with geometries that take advantage of magnification, to image smaller objects with higher position resolution.

The aim of this paper is to describe the performance of the detector, therefore, the practical example shown does not make a quantitative elemental analysis of the pigments. However, by applying corrections to the X-ray yield of each element, the self absorption and the detector and geometry efficiency as a function of the energy, it would be possible to estimate absolute concentrations in the different regions of the image.

The electronic system is very simple, with only five electronic channels. This has the advantage of cost effectiveness when the energies are above 6\,keV. Below these values, the signal-to-noise ratio becomes a problem and the loss of position resolution is inevitable. To solve this problem, the replacement of the acquisition system is foreseen. The new system will consist of discrete electronics, with 512 electronic channels readout via an ASIC. The elimination of the resistive chains will dramatically reduce the noise at a small cost, significantly improving the resolution of the images for low X-ray energies.

Other major improvements are planned for the detector itself, related to the improvement of the efficiency for lower and higher energies, by replacing the window by a thinner kapton foil and the cathode foil by a \SI{1.5}{\micro\m} thick aluminized polypropylene foil and by increasing the thickness of the drift region to around 20\,mm. Finally, a new \SI{100}{\micro\m} pinhole in a thin(\SI{200}{\micro\meter}) gold foil will also be tested, to make sure the improvements of this system in terms of position resolution will not be limited by geometrical features or by contamination of the spectra with fluorescences that do not occur in the sample.

\section*{Acknowledgments}

This work was supported by grants 2016/05282-2 and 2017/00426-9 from
Funda\c{c}\~ao de Amparo \`a Pesquisa do Estado de S\~ao Paulo, Brasil.

\bibliography{mybibfile}

\end{document}